\documentclass[aps,prl,twocolumn,superscriptaddress]{revtex4-2}
\usepackage{graphicx}
\usepackage{amsmath}
\usepackage{amssymb}
\usepackage{hyperref}
\usepackage{braket}
\usepackage{tikz}
\usepackage{quantikz}

\begin{document}

\title{Ground State Energy of He molecule Using a Four-Qubit Photonic Processor with the Variational Quantum Eigensolver}

\author{Badie Ghavami}
\email{ghavamiba@gmail.com}
\affiliation{ Iranian Center for Quantum Technologies (ICQT), Tehran, Iran}

\author{Forouzan Mirmasoudi}
\email{fmirmasoudi@uma.ac.ir}
\affiliation{ Iranian Center for Quantum Technologies (ICQT), Tehran, Iran}

\date{\today}

\begin{abstract}
To understand the properties and interactions of materials, and determining the ground state energies is one of the important challenges in quantum chemistry, materials science, and quantum mechanics, where quantum computing can play an important role for studying the properties of materials.  In this study, we have explored the quantum processor application to compute the He molecule ground state energy which utilizes the Variational Quantum Eigensolver (VQE) algorithm. In here, we have implemented VQE on a state-of-the-art quantum processor, optimizing a parameterized quantum circuit to minimize the energy expectation value of the He molecule's Hamiltonian on the four qubits processor.
The obtained results of this work show a significant improvement in accuracy compared to classical computational methods, such as Hartree-Fock and density functional theory, which demonstrate the compute potential of quantum algorithms in quantum many-body problems.
Thus, these results demonstrate the advantages of quantum computing in achieving high accuracy in simulations of molecular and material properties, and pave the way for future applications in more complex systems. This work highlights the potential of quantum processors in the fields of quantum chemistry, computational physics, and data science.
\end{abstract}

\maketitle

\section{Introduction}
The computation of ground state energies of molecules, and more broadly materials, is among the fundamental challenges of quantum chemistry, condensed matter, and materials science, with direct relevance to understanding their properties and interactions. For the study of molecular properties, the He molecule is used as a test system due to its simplicity and well-established theoretical and experimental results from standard computational methods such as Hartree-Fock and density functional theory (DFT) \cite{engel2011density, gross2013density, orio2009density, kryachko2014density,neugebauer2013density, cao2019quantum, sawayaquantum}. These traditional methods, however, are generally limited in their accuracy and computational cost, especially as the complexity of the systems under study increases. Recent advances in quantum computing have given a hopeful way out to surmount these challenges. Quantum processors, based on the foundations of quantum mechanics to perform calculations, have gained attention, unlike classical computer processors, which are unsolvable. Also, quantum algorithms such as the variational quantum eigensolver (VQE)\cite{liu2019variational} and the quantum approximate optimization algorithm (QAOA)\cite{farhi2014quantum} have shown great promise in the calculation of ground state energies and more accurate simulation of quantum many-body phenomena for small molecular systems. \cite{peruzzo2014variational, mcclean2016theory}.
The VQE in particular is a quantum-classical hybrid algorithm to minimize the energy of a quantum system by iteratively adjusted parameters in an ansatz circuit. The procedure has been successfully used on several molecular systems with efficiency in the electronic correlation calculation and molecular property understanding evidenced \cite{mcclean2016theory, kandala2017hardware}. Apart from this, the emergence of noise-resilient quantum algorithms and error mitigation techniques enhances the feasibility of near-term quantum hardware for practical quantum simulations \cite{temme2017error, endo2021hybrid,kandala2019error, cerezo2021variational, peruzzo2014variational}. When executed in a photonic quantum processor, VQE exploits the unique characteristics of photonic systems, such as long coherence times and the ability to perform some quantum operations efficiently \cite{peruzzo2014variational, sparrow2018simulating, wang2020integrated, knill2001scheme}. Photonic quantum processors implement photons as qubits \cite{flamini2018photonic}. They encode quantum information in properties of photons, such as polarization, path, or time bins. Photons are less susceptible to decoherence than other qubit technologies (e.g., superconducting qubits or trapped ions). 
In contrast to other quantum systems, photonic processors do not have to be cooled in order to function. Linear optics can implement some quantum operations, including beam splitters and phase shifters, with minimal efficiency. It is hard for photonic systems to achieve efficient nonlinear operations (e.g., two-qubit gates) and usually requires resources like ancillary photons or post-selection. Implementing the VQE on a photonic quantum processor means that specific consideration must be taken of the particular strengths and vulnerabilities of photonic platforms. Photonic quantum processors use photons as qubits, and they possess distinct advantages and limitations compared to other quantum computing platforms (such as superconducting qubits or trapped ions). \\
In this paper, we illustrate the application of a quantum processor to compute the ground-state energy of the He molecule. Through VQE, we contrast quantum computations with conventional classical calculations and illustrate the advantage of quantum computing in achieving greater precision and efficiency for quantum chemical questions.\\
\section{Methodology}
To describe the quantum algorithms, we are employing processor details of quantum, and  computation techniques applied to calculate ground-state energy of He molecule. The Hamiltonian of He molecule is the sum of the kinetic energy and potential energy of the electrons and their coupling. The Hamiltonian expressed as Pauli operators can be applied on a quantum computer.  The Hamiltonian for the He atom is given by \cite{frs1947calculation,szabo1996modern}:
\begin{equation}
\mathcal{\hat{H}}=-\frac{\hbar^2}{2m}\nabla^2 +\mathcal{V}(r)
\end{equation}
where $\hbar$, $m$, and $\mathcal{V}(r)$ are the reduced Planck constant, the electron mass,  and 
the Coulomb potential between the electrons and the nuclei respectively, in which the Coulomb potential is given by:
\begin{equation}
\mathcal{V}(r)=-\frac{2e^2}{4\pi\epsilon_0 r_1}-\frac{2e^2}{4\pi\epsilon_0 r_2}+\frac{e^2}{4\pi\epsilon_0 r_{12}}
\end{equation}
where $r_1$ and $r_2$ represent the distance of the electrons from the nucleus, and $r_{12}$
is the distance between the two electrons.
This Hamiltonian provided us with the first-quantized Hamiltonian, which outlined the system in terms of electron positions and momenta. This Hamiltonian must be translated into second-quantized form using creation and annihilation operators in order to utilize this Hamiltonian for quantum computing. This allows us to express the Hamiltonian through Pauli operators, which can be utilized within a quantum computer. In the process that is given below, we can see how to map the first-quantized Hamiltonian into an expression of the second quantized and further on Pauli operators using the Jordan-Wigner transformation.
The second-quantized Hamiltonian of the He atom is \cite{bruus2004many}:
\begin{equation}\label{Hamiltonian}
\hat{H} = \sum_{pq} h_{pq} a_p^\dagger a_q + \frac{1}{2} \sum_{pqrs} V_{pqrs} a_p^\dagger a_q^\dagger a_r a_s,
\end{equation}
where  \(a_p^\dagger\) and \(a_p\) are the creation and annihilation operators for the fermionic modes, and
\(h_{pq}\) are the one-electron integrals (kinetic energy and electron-nucleus attraction) which are given by:
\begin{equation}\label{hopping}
h_{pq} = \int \phi_p^*(r) \left( -\frac{\hbar^2}{2m} \nabla^2 - \frac{2e^2}{4\pi\epsilon_0 r} \right) \phi_q(r) \, d^3r.
\end{equation}
This integral represents the kinetic energy and electron-nucleus attraction for the electrons in orbitals $p$ and $q$. The two-electron integrals $V_{pqrs}$ (electron-electron repulsion) are given by:
\begin{equation}\label{Vq}
V_{pqrs} = \int \phi_p^*(r_1) \phi_q^*(r_2) \left( \frac{e^2}{4\pi\epsilon_0 r_{12}} \right) \phi_r(r_1) \phi_s(r_2) \, d^3r_1 \, d^3r_2.
\end{equation}
This integral represents the electron-electron repulsion between electrons in orbitals $p$, $q$, $r$, and $s$.\\
The Jordan-Wigner transformation maps fermionic operators to spin (Pauli) operators \cite{batista2001generalized}. For a system with \(N\) fermionic modes, the transformation is given by:
\begin{align}\label{ap}
a_p^\dagger = \prod_{k=1}^{p-1} \sigma_k^z \cdot \sigma_p^+, \\
a_p = \prod_{k=1}^{p-1} \sigma_k^z \cdot \sigma_p^-,
\end{align}
where \(\sigma_k^z\)  is the Pauli-Z operator acting on the \(k\)-th qubit, \(\sigma_p^+ = \frac{1}{2} (\sigma_p^x + i \sigma_p^y)\) and \(\sigma_p^- = \frac{1}{2} (\sigma_p^x - i \sigma_p^y)\) are the spin raising and lowering operators.\\
Let’s apply the Jordan-Wigner transformation to the second-quantized Hamiltonian for the He atom. The one-electron term \(a_p^\dagger a_q\) becomes:
\begin{align}\label{apaq}
a_p^\dagger a_q = \left( \prod_{k=1}^{p-1} \sigma_k^z \cdot \sigma_p^+ \right) \left( \prod_{k=1}^{q-1} \sigma_k^z \cdot \sigma_q^- \right).
\end{align}
For the He atom, we have two orbitals (\(p, q = 1, 2\)), so the one-electron terms are:
\begin{align}\label{a1a1}
a_1^\dagger a_1 = \sigma_1^+ \sigma_1^- = \frac{1}{2} (I - \sigma_1^z), \\
a_2^\dagger a_2 = \sigma_1^z \sigma_2^+ \sigma_2^- = \frac{1}{2} \sigma_1^z (I - \sigma_2^z).
\end{align}
The two-electron term \(a_p^\dagger a_q^\dagger a_r a_s\) becomes:
\begin{align}\label{apaq}
a_p^\dagger a_q^\dagger a_r a_s = ( \prod_{k=1}^{p-1} \sigma_k^z \cdot \sigma_p^+) ( \prod_{k=1}^{q-1} \sigma_k^z \cdot \sigma_q^+ )  \nonumber \\ ( \prod_{k=1}^{r-1} \sigma_k^z \cdot \sigma_r^- ) ( \prod_{k=1}^{s-1} \sigma_k^z \cdot \sigma_s^- ).
\end{align}
For the He atom, the only relevant two-electron term is \(a_1^\dagger a_2^\dagger a_1 a_2\), which becomes:
\begin{align}\label{aa1}
a_1^\dagger a_2^\dagger a_1 a_2 = \sigma_1^+ \sigma_1^z \sigma_2^+ \sigma_2^- = \frac{1}{4} (I - \sigma_1^z) (I - \sigma_2^z).
\end{align}
Combining the one-electron and two-electron terms, the Hamiltonian for the He atom after the Jordan-Wigner transformation can be obtained.
\begin{align}
\hat{H}& = \sum_{pq} h_{pq} ( \prod_{k=1}^{p-1} \sigma_k^z \cdot \sigma_p^+ ) ( \prod_{k=1}^{q-1} \sigma_k^z \cdot \sigma_q^- ) \nonumber \\
&+ \frac{1}{2} \sum_{pqrs} V_{pqrs} ( \prod_{k=1}^{p-1} \sigma_k^z \cdot \sigma_p^+) ( \prod_{k=1}^{q-1} \sigma_k^z \cdot \sigma_q^+ ) \nonumber \\
& \times( \prod_{k=1}^{r-1} \sigma_k^z \cdot \sigma_r^- ) ( \prod_{k=1}^{s-1} \sigma_k^z \cdot \sigma_s^- ).
\end{align}
%
%
%
%
%
%
 For improved calculation accuracy, post-Hartree-Fock theories such as Perturbation Theory and Configuration Interaction may be employed \cite{custodio2018grid}. Such theories incorporate electron correlation effects that the Hartree-Fock method does not.\\
 
 \section{implementation and result}
We are using the VQE algorithm is used to find the ground state energy of a quantum system, in which the algorithm uses a parameterized quantum circuit (ansatz) and a classical optimizer to minimize the expected energy, $
\mathcal{E}(\theta)=\langle\psi(\theta)\vert\mathcal{\hat{H}}\vert\psi(\theta)\rangle
$, in which $\theta$ is the parameter of the quantum circuit, and $\hat{H}$ is the Hamiltonian of
the system. The goal of the VQE algorithm is to find the optimal parameters $\theta^*$ that minimize the expected energy: $\theta^*=arg \ \min_\theta \mathcal{E}(\theta)$.
To calculate the ground state energy of He, we use the following quantum circuits: firstly, 
we define a quantum circuit with 4 qubits, then the CNOT gates are applied between qubits 0 and 1, and between qubits 2 and 3, and finally a more complex Mach-Zehnder interferometer (MZI) is added between qubits (1 , 2), and (2 , 3).\\
We have used Hadamard gates for create superposition, and CNOT gate with qubit 0 (2) as the control and qubit 1 (3) as the target, creating entanglement between these qubits. The Mach-Zehnder interferometers have added to create complex interferences between the qubits and manipulate their phases, which have represented with directional couplers and a gate $R_z$ (phase shifter).\\
The unitary matrix (Directional Coupler) $U$ of a quantum circuit represents the general transformation applied to the quantum state by the circuit that each quantum gate in the circuit contributes to this unitary transformation, and the combined effect of all gates can be described by a single unitary matrix. It can be used to perform any rotation or unitary transformation on a single qubit. Here, the $U$ applies a unitary transformation to a qubit and is represented by the following matrix\ref{unitary}:
\begin{equation}\label{unitary}
U(\theta, \phi, \lambda)= 
\begin{pmatrix}
	\cos(\theta/2) & -e^{-i\lambda}\sin(\theta/2)\\
	e^{i\phi}\sin(\theta/2) & e^{i(\phi+\lambda)}\cos(\theta/2)\\
\end{pmatrix} 
\end{equation}
where  $\theta= \arccos( R)$ controls the rotation around the y-axis ( R represents
the coupling ratio of the directional coupler),  $\phi$,  and $\lambda$ control the rotations
around the z-axis. In quantum circuit of Fig.\ref{fig:quantum_circuit}, $\phi = \pi/2$, $\lambda = 0$, that these gates apply
a rotation around the y-axis of the Bloch sphere determined by the values of $R_1$
and $R_2$.\\
The $R_z$ gate is a single-qubit rotation gate that applies a rotation about the z-axis of the Bloch sphere. The matrix representation of the $R_z$ gate is given by:
\begin{equation}
R_z(\beta)=\begin{pmatrix}
	e^{-i\beta/2} & 0\\
	0 & e^{i\beta/2}\\
\end{pmatrix}
\end{equation}
where $\beta$ is the rotation angle. In the quantum circuit, the $R_z$ gate is used to
adjust the phase of the qubit which is particularly useful in creating interference
patterns and controlling the state of the qubits. In the context of the Mach-Zehnder interferometer, $R_z$ gates are applied to specific qubits to achieve the desired phase shifts. Schematic structure of the optical chip manufactured by Mach-Zehnder interferometer is shown in Fig.\ref{fig:quantum_circuit}.
\begin{figure}[h]
  \centering
   \scalebox{0.477}
{  
    \begin{quantikz}
    \lstick{$q_0$} & \gate[style={fill=blue!20}]{H} & \ctrl{1} & \qw      & \qw      & \qw & \qw      & \qw   & \qw  & \qw & \qw & \qw & \qw & \qw    & \qw \\
    \lstick{$q_1$} & \qw      & \targ{}  & \gate[style={fill=red!20}]{U(\pi/2)} & \ctrl{1} & \qw & \ctrl{1} & \gate[style={fill=red!20}]{U(\pi/2)} & \qw      & \qw & \qw & \qw  & \qw  & \qw  & \qw \\
    \lstick{$q_2$} & \gate[style={fill=blue!20}]{H}   & \ctrl{1}    & \qw      & \targ{}  & \gate[style={fill=green!20}]{R_z(\pi/2)} & \targ{}
    & \gate[style={fill=green!20}]{R_z(0)} & \qw &\gate[style={fill=red!20}]{U(\pi/2)}   & \ctrl{1} &\qw  & \ctrl{1}    & \gate[style={fill=red!20}]{U(\pi/2)} & \qw   \\
    \lstick{$q_3$} & \qw  & \targ{}    & \qw   &\qw &\qw   & \qw      & \qw      & \qw & \qw    & \targ{}  & \gate[style={fill=green!20}]{R_z(\pi/2)} & \targ{}  & \gate[style={fill=green!20}]{R_z(0)}  &\qw
    \end{quantikz}
    }
    \caption{Schematic image of the four qubit chip waveguide structure used for
simulation with consider $U=U(\pi/2, \pi/2, \pi/2)$ .}
    \label{fig:quantum_circuit}
\end{figure}
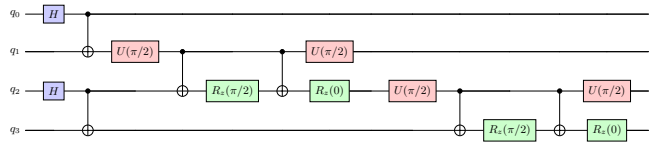
\begin{figure*}[ht]
\centering
\includegraphics[scale=0.31]{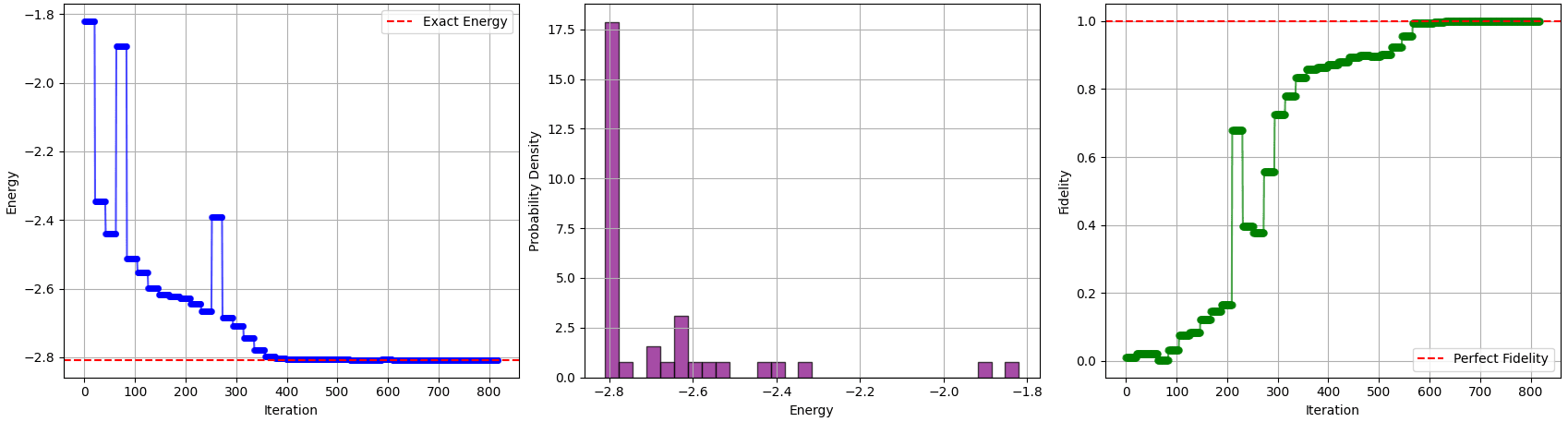}
\caption{VQE convergence for He Molecule and probability density for measured energies, fidelity during optimization and by using quantum photonic processing from left to right, respectivelity.}
\label{fig2}
\end{figure*}
 We numerically investigated the influence of power splitting ratios and phase shifts in calculation of ground state energy of He using an open-source framework for quantum computing (Qiskit)\cite{singh2021quantum, aleksandrowicz2019qiskit}, and it is shown in Fig.\ref{fig2}. In case of an ideal
chip $R_1  =  R_2  =  0.5 $ and $\phi=\pi/2$, $\lambda=\pi/2$.
The x-axis represents the number of iterations, and the y-axis represents the energy values obtained during the optimization process. As the iterations increase, the energy values should ideally
converge towards the ground state energy of the He molecule. The green line indicates how the energy values change over the iterations, showing the
progress of the optimization.
The second plot in Fig.\ref{fig2} is a histogram showing the distribution of the measured energy values throughout the VQE optimization process. The x-axis
represents the energy values, and the y-axis represents the frequency of these values. This plot helps visualize how often certain energy values were obtained, giving an idea of the stability and variability of the optimization process.
In quantum computing, fidelity is a measure of the closeness between two quantum states. It quantifies how well a quantum system can prepare, manipulate, and measure states compared to an ideal target state. High fidelity is essential for reliable quantum computation, as it ensures that the results of quantum operations are accurate and trustworthy.
For two pure quantum states \(|\psi\rangle\) and \(|\phi\rangle\), the fidelity \(F\) is defined as the squared overlap between the states:
\begin{equation}
F(|\psi\rangle, |\phi\rangle) = |\langle \psi | \phi \rangle|^2
\end{equation}
Where, \(|\psi\rangle\) represents the state produced by the quantum system, and \(|\phi\rangle\) is the target state. 
A fidelity of 1 indicates that the states are identical, while a fidelity of 0 means they are orthogonal.
Fidelity plays a critical role in evaluating the performance of quantum processors. For example, in photonic quantum computing, fidelity is used to assess the quality of quantum state preparation, gate operations, and measurements. In variational quantum algorithms (VQA)\cite{cerezo2021variational} like the VQE, fidelity measures how closely the final state produced by the quantum circuit approximates the target ground state of a Hamiltonian.\\
Achieving high fidelity in quantum systems is challenging due to factors such as noise, decoherence, and imperfections in quantum hardware. In photonic quantum processors, for instance, photon loss and imperfect optical components can significantly reduce fidelity. Addressing these challenges is crucial for scaling quantum technologies and achieving practical quantum advantage.

In this work, we use fidelity to evaluate the performance of our photonic quantum processor in simulating the ground state of a helium-like system (it's shown Fig\ref{fig2}). By tracking the fidelity during the optimization process, we assess how well our quantum circuit approximates the target state and identify areas for improvement.
Fidelity is a key metric in quantum computing, quantifying the closeness between two quantum states. For pure states \(|\psi\rangle\) and \(|\phi\rangle\), fidelity is defined as \(F(|\psi\rangle, |\phi\rangle) = |\langle \psi | \phi \rangle|^2\), where a value of 1 indicates perfect overlap and 0 indicates orthogonality. In quantum processors, fidelity is used to evaluate the accuracy of state preparation, gate operations, and measurements. Achieving high fidelity is challenging due to noise, decoherence, and hardware imperfections, particularly in photonic systems where photon loss and optical component errors can degrade performance. In this work, we use fidelity to assess the performance of our photonic quantum processor in simulating the ground state of a helium-like system, providing insights into the quality of our quantum circuit and the effectiveness of our optimization process. Table~\ref{tab:comparison} provides a comparative analysis of our work with existing studies on simulating helium-like quantum systems. While prior efforts focused on superconducting \cite{houck2012chip} or trapped-ion qubits \cite{monroe2021programmable}, our methodology leverages a photonic quantum processor, enabling unique energy distribution analyses alongside standard fidelity metrics. The custom ansatz (RealAmplitudes + photonic circuits) further distinguishes our approach, as highlighted in Column 3. This comparison demonstrates how our hardware and algorithmic innovations address scalability challenges in quantum simulations, a limitation noted in earlier works.

\begin{table*}[t!]
\centering
\caption{Comparison He ground state energy which previous studies with this work}
\begin{tabular}{|l|l|l|}
\hline
\textbf{Aspect} & \textbf{Previous Studies} & \textbf{This Work} \\ \hline
\textbf{Problem} & 
\begin{tabular}[c]{@{}l@{}}
Simulating helium-like systems on\\ superconducting/trapped-ion qubits.
\end{tabular} & 
\begin{tabular}[c]{@{}l@{}}
Simulating helium-like systems on\\ a photonic quantum processor.
\end{tabular} \\ \hline

\textbf{Methodology} & 
\begin{tabular}[c]{@{}l@{}}
VQE with standard ansatzes\\ (e.g., UCC, Hardware-Efficient).
\end{tabular} & 
\begin{tabular}[c]{@{}l@{}}
VQE with a custom ansatz\\ (RealAmplitudes + photonic circuit).
\end{tabular} \\ \hline

\textbf{Results} & 
\begin{tabular}[c]{@{}l@{}}
Fidelity and energy convergence\\ plots are common.
\end{tabular} & 
\begin{tabular}[c]{@{}l@{}}
Fidelity, energy convergence, and\\ energy distribution analysis.
\end{tabular} \\ \hline
\textbf{Hardware} & 
\begin{tabular}[c]{@{}l@{}}
Superconducting or trapped-ion qubits.
\end{tabular} & 
\begin{tabular}[c]{@{}l@{}}
Photonic quantum processor.
\end{tabular} \\ \hline
\end{tabular}
\label{tab:comparison}
\end{table*}
\subsection{Computation of Matrix Permanents} 
This section presents the theoretical foundation and computational methods for calculating matrix permanents, which are integral to simulating quantum photonic processors. We provide an algorithmic approach using Ryser's method for efficient permanent computation, and we demonstrate its application in determining the probabilities of various Fock state outputs in a quantum photonic circuit.  The matrix permanent is a fundamental concept in combinatorial mathematics and quantum computing. Unlike the determinant, the permanent of a matrix does not involve alternating signs, making its computation more complex and computationally intensive. The permanent of an $n \times n$ matrix $A$ is defined as \cite{tillmann2013experimental}:
 \begin{equation} 
\text{perm}(A) = \sum_{\sigma \in S_n} \prod_{i=1}^n a_{i,\sigma(i)} 
\end{equation}
 where $S_n$ is the set of all permutations of $\{1, 2, \ldots, n\}$, and $a_{i, \sigma(i)}$ denotes the elements of the matrix $A$ corresponding to the permutation $\sigma$. Ryser's algorithm offers an efficient way to compute the permanent of a matrix by reducing the complexity compared to the naive approach. The permanent can be computed using the following steps: 1. Initialize the sum to zero. 2. Iterate over all $2^n$ subsets of the set $\{1, 2, \ldots, n\}$. 3. For each subset, compute the product of the sums of selected rows, adjusted by the parity of the subset size. 4. Sum the results to obtain the permanent. 
In quantum photonic processors, matrix permanents play a crucial role in calculating the probabilities of various Fock state outputs. Given a unitary matrix $U$ representing the quantum circuit, the probability $P(\text{out}|\text{in})$ of measuring a specific output Fock state from a given input Fock state is given by:
\begin{equation}
P(\text{out}|\text{in}) = \frac{|\text{perm}(U_{\text{in}, \text{out}})|^2}{\prod_{i=1}^n \text{in}_i! \prod_{j=1}^n \text{out}_j!}
\end{equation}
where $U_{\text{in}, \text{out}}$ is the submatrix of $U$ indexed by the input and output Fock states, and $\text{in}_i$ and $\text{out}_j$ are the occupation numbers of the input and output Fock states, respectively.
The computation of matrix permanents is essential for simulating quantum photonic processors. Ryser's algorithm provides an efficient method for calculating these permanents, making it feasible to determine the probabilities of various Fock state outputs in quantum circuits. This work highlights the theoretical foundations and practical implementations necessary for advanced quantum computing research.
\begin{figure*}[ht]
\centering
\includegraphics[scale=0.5]{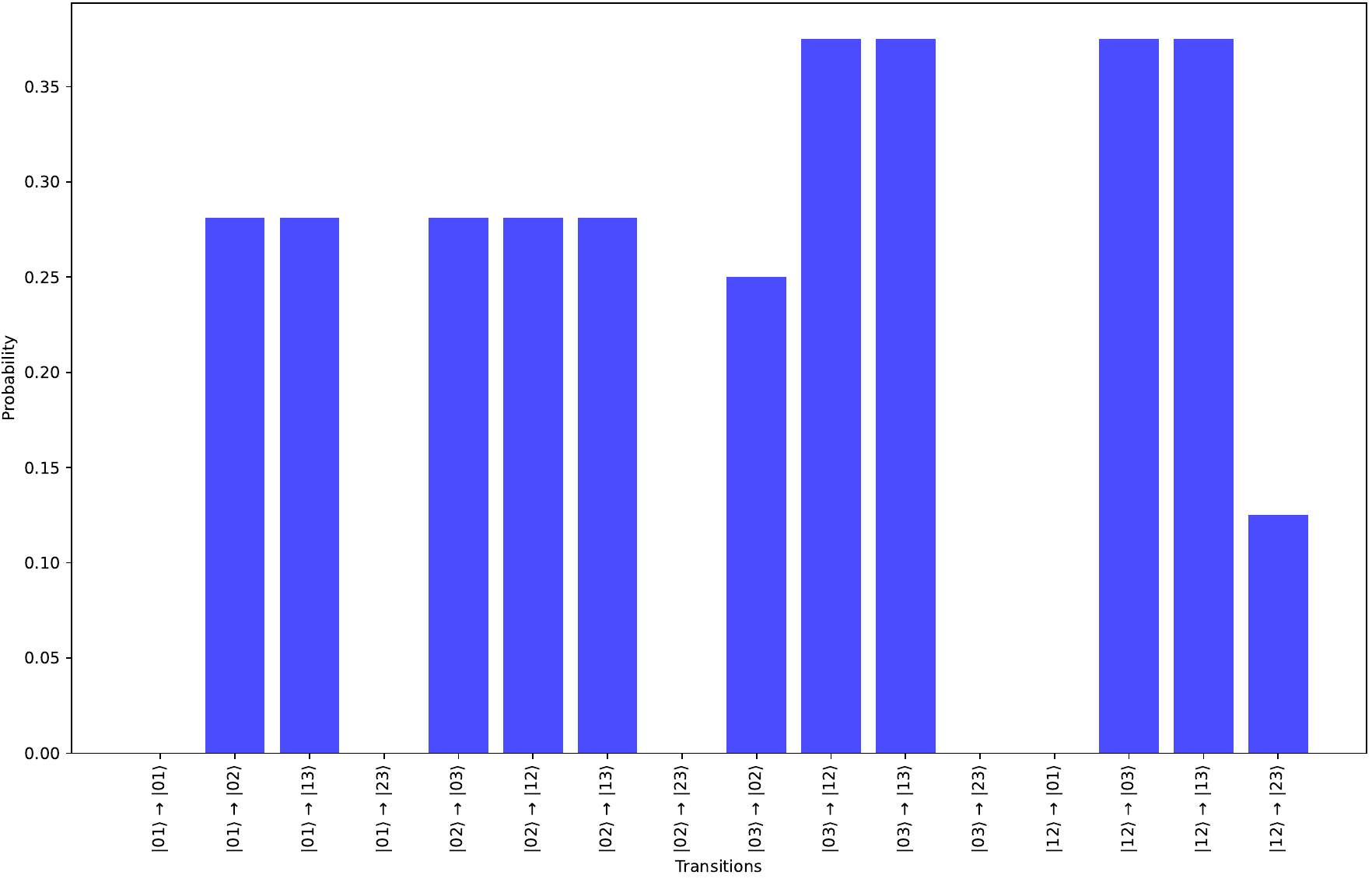}
\caption{Bar Plot of Measuring Output Fock States}
\label{fig3}
\end{figure*}
Fock states, also known as number states, describe a definite number of particles in a given mode. They are fundamental in quantum optics and quantum information theory.
For systems with multiple modes, the Fock state notation extends to describe the number of particles in each mode. If there are \(m\) modes, the Fock state is represented as:
A Fock state with \(n_1\) particles in mode 1, \(n_2\) particles in mode 2, up to \(n_m\) particles in mode \(m\):
\begin{align}
  |n_1, n_2, \ldots, n_m\rangle
  \end{align}
Examples of multi-mode Fock states include: a state with 1 particle in the first mode and 0 particles in the second mode:
\(
  |1, 0\rangle
  \).
 A state with 2 particles in the first mode and 3 particles in the second mode:
\(
  |2, 3\rangle
  \).
These Fock states are crucial in describing the quantum states of particles in various modes and are widely used in research involving quantum mechanics and quantum optics.
The probabilities of different transitions between Fock states for  photonic quantum processor  is visualized in Fig.(\ref{fig3}). Each bar represents a specific transition from an input Fock state to an output Fock state, with the height of the bar indicating the probability of that transition occurring.
The chart indicates the likelihood of transitions between different Fock states when photons propagate through the optical chip described by the unitary matrix $U$. For example, a bar labeled $\left|01\right\rangle \rightarrow \left|02\right\rangle$ with a height indicating a probability of approximately 0.3 suggests that there is a 30\% chance of the photons transitioning from the state $\left|01\right\rangle$ to $\left|02\right\rangle$. Some transitions, such as 
$\left| 03 \right\rangle \rightarrow \left| 02 \right\rangle$, 
$\left| 03 \right\rangle \rightarrow \left| 12 \right\rangle$, 
$\left| 03 \right\rangle \rightarrow \left| 13 \right\rangle$, 
$\left| 03 \right\rangle \rightarrow \left| 23 \right\rangle$, 
have different probabilities. This demonstrates the flexibility and capabilities of the processor in processing photons.
Therefore,
the processor shows a consistent performance in most transitions, ensuring reliable quantum state processing.
Also, the variation in probabilities for certain transitions highlights the processor's ability to handle different quantum states with varying efficiency.
The transitions and their corresponding probabilities provide insights into the behavior of the photonic quantum processor and the interaction between photons within the system.
This type of visualization is crucial in quantum optics and quantum information theory, as it helps researchers understand and analyze the quantum state transitions and the underlying probabilities in a quantum system.

\section{Conclusion}
Using quantum computing and the $VQE$ algorithm on a photonic quantum
processor, the ground state energy of the He molecule was computed. 
The main purpose  of this work was to demonstrate the capability of quantum computation methods than classical methods such as Hartree-Fock (HF) and density functional theory (DFT) in determining ground state energies. To this end,  we have optimized a parameterized quantum circuit to minimize the energy expectation value of the He Hamiltonian.
This results show strong agreement with theoretical and experimental benchmarks.
Moreover, the calculation of fidelity, energy convergence, and transition probabilities between Fock states, have demonstrated  the robustness of our approach and its alignment with theoretical benchmarks.
The results pave the way towards large-scale simulations of larger molecular systems, offering a promising direction to tackle quantum many-body issues in molecular physics and materials science. It seems  that our finding may sound interesting in quantum computing.

\section{acknowledgments}
We permit the utilization of AI-enabled tools (e.g. Grammarly, DeepSeek) for proofreading and improving the organization and clarity of this manuscript. However, all scientific content, analysis, and conclusions are within the responsibility of the authors.

\bibliography{references}

\end{document}